\documentclass[journal=jctcce,manuscript=article]{achemso}
\setkeys{acs}{articletitle = true}

\usepackage[version=3]{mhchem} 
\usepackage{multirow}

\author{Hyeondeok Shin}
\affiliation{Leadership Computing Facility, Argonne National Laboratory, Argonne, IL 60439, United States}
\author{Jeongnim Kim}
\affiliation{Intel Corporation, Hillsboro, Oregon 97124, United States}
\author{Hoonkyung Lee}
\affiliation{Department of Physics, Konkuk University, Seoul 05029, Korea}
\author{Olle Heinonen}
\affiliation{Material Science Division, Argonne National Laboratory, Argonne, IL 60439, United States}
\author{Anouar Benali}
\email{benali@anl.gov}
\phone{+1-630-252-0058}
\fax{+1-630-252-0374}
\affiliation{Leadership Computing Facility, Argonne National Laboratory, Argonne, IL 60439, United States}
\author{Yongkyung Kwon}
\email{ykwon@konkuk.ac.kr}
\phone{+82-2-450-3410}
\fax{+82-2-3436-5382}
\affiliation{Department of Physics, Konkuk University, Seoul 05029, Korea}

\title{The Nature of Interlayer Binding and Stacking of $sp$-$sp^{2}$ Hybridized Carbon Layers: A Quantum Monte Carlo Study}


\begin{document}




\begin{abstract}
$\alpha$-graphyne is a two-dimensional sheet of $sp$-$sp^2$ hybridized carbon atoms in a honeycomb lattice. While the geometrical structure is similar to that of graphene, the hybridized triple bonds give rise to electronic structure that is different from that of graphene.
Similar to graphene, $\alpha$-graphyne can be stacked in bilayers with two stable configurations, but the different stackings have very different electronic structures: one is predicted to have gapless parabolic bands and the other a tunable band gap which is attractive for applications. In order to realize 
applications, it is crucial to understand which stacking is more stable. This is difficult to model, as the stability is a result of weak interlayer van der Waals interactions which are not well captured by density functional theory (DFT). 
We have used quantum Monte Carlo simulations that accurately include van der Waals interactions to calculate the interlayer binding energy of bilayer graphyne and to determine its most stable stacking mode. Our results show that 
interlayer bindings of $sp$- and $sp^{2}$-bonded 
carbon networks are significantly underestimated 
in a Kohn-Sham DFT approach, even with 
an exchange-correlation potential corrected to include, in some approximation, van der Waals interactions. 
Finally, our quantum Monte Carlo calculations reveal
that the interlayer binding energy difference between the two stacking modes is only 0.9(4)~meV/atom.
From this we conclude that the two stable stacking modes of bilayer $\alpha$-graphyne
are almost degenerate with each other, and both will occur with about the same probability at room temperature unless there is a synthesis path that prefers one stacking over the other.
\end{abstract}
\newpage
	
\section*{Introduction}
\label{sec:introduction}
Low-dimensional carbon allotropes have been extensively studied because of their exotic electronic and mechanical properties. 
In particular, a single sheet of graphite, graphene, has a great potential for future 
nanoelectronic and spintronic devices 
based on its massless Dirac fermion physics and anomalous quantum Hall effects.~\cite{novoselov05,zhang05,geim07,geim09}
It has also been shown that bilayer graphene (two layers of graphene sheets) possesses significantly different electronic properties than those of a single-layer graphene, with the difference depending on the stacking mode of the bilayer.~\cite{park15} For instance, a perfectly parallel-aligned stacking mode along the $z$ direction (AA) shows nearly metallic properties in its electronic band structure. In contrast, the Bernal stacking mode (AB), in which  
half of carbon atoms in the upper graphene layer are located at the hexagon centers of the lower layer,
exhibits a tunable band gaps when an external electronic field is applied normal to the surface.~\cite{ohta06,castro07,oostinga08}
Because the most stable mode of graphite in nature is known to be the Bernal mode (ABA), it was trivial to identify the most energetically stable mode of a bilayer graphene as the Bernal AB mode.~\cite{yan11} 
However, quantitative measurement of the energetics of interlayer bindings has proceeded slowly compared to studies of other electronic properties, mainly because the interlayer interaction between each single-layer graphene are completely dominated by weak intermolecular van der Waals interactions arising from electron correlations between the hybrid $p_{z}$ orbitals in each layer.

In addition to single- and bi-layer graphene sheets, other low-dimensional carbon allotropes have attracted a great deal of attention because of their unique electronic properties related to their different structural complexes. 
An $sp$-$sp^{2}$ hybridized graphyne structure, the existence of which was predicted in the past few decades,~\cite{baughman87,coluci03,coluci04} has been expected to possess better electronic properties than graphene, with applications in future energy storage devices.~\cite{zhang11,srinivasu12,malko12,kim12,hwang12,chen13,hwang13}
Among several predicted forms of graphynes, $\alpha$-graphyne has a honeycomb structure with a unit cell larger than that of graphene.
Bilayers of $\alpha$-graphyne have been predicted~\cite{leenaerts13} to   
stabilize in two different stacking modes, AB and Ab modes, amongst six possible configurations (see Figure 2 of Ref.~\cite{leenaerts13}).
Because an AB-stacked bilayer $\alpha$-graphyne and AB bilayer graphene are aligned with the same shift between the unit cells of each upper and lower layer, an AB stacking $\alpha$-graphyne was naturally predicted to possess similar electronic properties as an AB bilayer graphene with parabolic bands touching at the K-point in the Brillouin zone~\cite{leenaerts13}.
In contrast, the Ab mode was predicted~\cite{leenaerts13} to possess split Dirac cones at the Fermi level near the K-point, with an applied electric field normal to the surface opening a gap at the Dirac points. This property makes the Ab-stacking very attractive for potential applications; in order to realize applications it is therefore important to accurately know the binding energy of the different stackings, in particular if the Ab stacking is more stable than the AB stacking.

Theoretical predictions of the electronic properties and binding energies of bilayer 
structures such as bilayer graphene, have been primarily conducted using DFT-based first-principle calculations.~\cite{ferrari06,sahu08,yang09,gava09,birowska11,mapasha12} 
DFT usually describes electronic properties with an acceptable level of accuracy and predictive powers. However, its usual implementations do not include descriptions of dispersive forces, the origin of van der Waals (vdW) forces, and without some corrections for vdW forces, DFT fails to describe the intermolecular interaction involved in the binding of two graphene layers.

There have been various attempts to calculate interlayer binding energies of a bilayer graphene using vdW-corrected DFT, but the results proved to be strongly dependent on the  choice of the vdW correction and the exchange-correlation (XC) functional.
For example, with an empirical vdW correction of Grimme (DFT-D),~\cite{grimme04,grimme06,grimme10} binding energies for the AA and AB stacking mode of a bilayer graphene were found to be 31.1~meV/atom and 50.6~meV/atom, respectively, while using a vdW-correction based on self-consistent non-local electron correlation (vdW-DF),~\cite{dion04,thonhauser07} resulted in binding energies of 10.4~meV/atom and 29.3~meV/atom, respectively.~\cite{lebedeva11} Because of the meV/atom scale of the binding energies of bilayer graphene, it is extremely challenging to accurately assess the binding energies either with theoretical or experimental methods.

For graphyne, even weaker binding energies than for bilayer graphenes were predicted by DFT, 
which was understood to be due to larger surface area of graphyne.
However, these calculations failed to conclusively predict the most stable stacking mode 
because the lower-energy stacking mode was strongly dependent on the type of the vdW-correction to the XC functional.~\cite{leenaerts13} 
The weak binding energies and the high uncertainty in DFT estimates of the relative stabilities of different stackings make it imperative to use a more accurate method to conclusively calculate the interlayer binding energies.
In contrast to DFT methods that are based on the electronic density as the independent variable, quantum Monte Carlo (QMC) methods work explicitly with the many-body wavefunction and the full Hamiltonian of the electronic system, and therefore rigorously include dispersion forces, such as vdW, and dynamic correlations. Also, no damping-function corrections need to be added to eliminate double-counting of interactions at small bonding distances as is the case in DFT where both exchange-correlation functionals and vdW functionals will contribute to the energy. Consequently, QMC has proven to be the method of choice to accurately describe electronic properties of many-body systems, and hence has been widely used to study vdW-dominated  systems.~\cite{drummond07,spanu09,shulenburger13,benali14,ganesh14,hsing14,mostaani15,shulenburger15}
More recently, QMC, or more accurately diffusion Monte Carlo (DMC), was used to study the interlayer binding energies of a bilayer graphene.~\cite{mostaani15} This study showed that previous vdW-corrected DFT calculations strongly {\em overestimate} interlayer binding energies both for AA and AB bilayer graphenes, and pointed to a path for better corrections of the dispersive forces in DFT functionals. 

In this paper, we use DMC to investigate the stability of the AB and Ab bilayer graphynes.
The comparison of our DMC results to those for AB bilayer graphene
reveals that bilayer graphyne have larger interlayer binding energies than bilayer graphene.
By analyzing the bilayer dissociation energy curve and the charge density of the systems, we
attribute the relative stability of bilayer graphyne over bilayer graphene 
to the contribution of interlayer covalent bonds in graphyne 
to the total energy of the system.
We then use the same method of analysis of dissociation energies and charge density to investigate various vdW-corrected DFT functionals, and our analysis points to the reasons why they fail to reproduce DMC results. The global analysis of the DFT results allows us to formulate a strategy for choosing the right vdW-corrected functionals in studies related to layered carbon systems. 

\section*{Methods}
\label{sec:methods}
Our study was carried out within the fixed node Diffusion Monte Carlo method~\cite{foulkes01,Reynolds1982} as implemented in the QMCPACK code.~\cite{kim12-1} We used a single Slater-Jastrow trial wavefunction with one and two-body variational Jastrow factors (ion-electron and electron-electron) to describe with sufficient accuracy the electronic correlation.  In order to simulate the layered systems, we used a supercell with  periodic boundary conditions in the $xy$ plane and vacuum in the non-periodic $z$ direction. The size of the vacuum was converged using DMC to a value of 40 \AA.
 Within each layer, the geometry was fully optimized using the Vienna {\it{Ab initio}} Simulation Package (VASP) until the atomic forces were less than 0.01 eV/$\text{\AA}$.
 The antisymmetric fermionic part of the wavefunction was calculated within the DFT framework using a plane-wave  basis set of  300~Ry cutoff and a $12\times12\times1$ Monkhorst-Pack $k$-point grids~\cite{monkhorst76} and to generate the single particle orbitals. The self-consistent DFT calculations were performed with the Perdew-Burke-Ernzerhof (PBE) XC functional~\cite{perdew96} as implemented in the QUANTUM ESPRESSO package.~\cite{giannozzi09} 
 In order to reduce computational costs, all calculations used pseudopotentials proposed by Burkatzki, Filippi, and Dolg (BFD)~\cite{burkatzki07,burkatzki08}, the accuracy of which was demonstrated in a previous work for similar systems.~\cite{shin14}
Results were converged with a time step of 0.005 Ha$^{-1}$ and the $T$-move approximation was employed to localize the employed BFD non-local pseudopotential in the effective Hamiltonian.~\cite{casula06}
Finite-size effects were controlled by applying twist-averaged boundary condition (TABC) (one body effects~\cite{lin01}) and extrapolating the supercell to infinite size (two body effects). 
\begin{figure}[t]
 \includegraphics[width=3.0in]{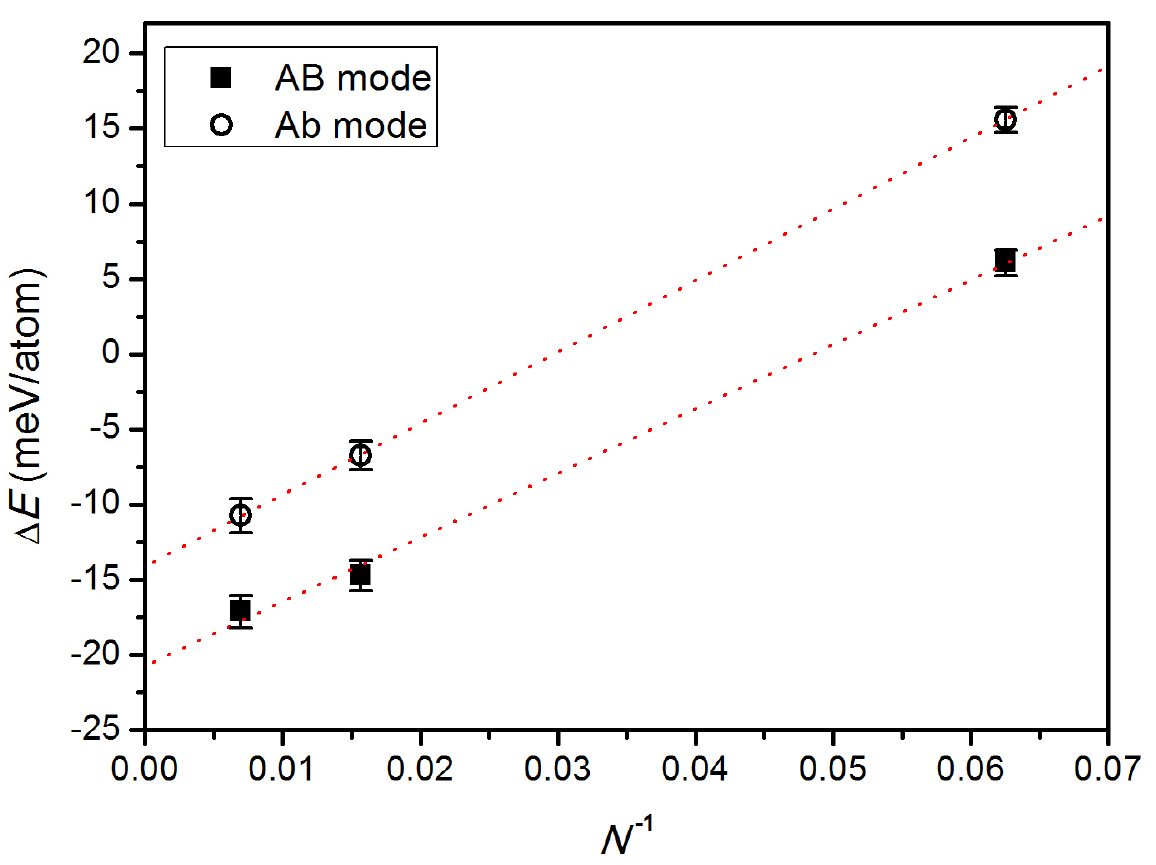}
 \caption{QMC interlayer binding energies of AB and Ab stacking modes for a bilayer $\alpha$-graphyne at an interlayer distance of 3.00 $\text{\AA}$ as function of inverse number of atoms per supercell. The dotted lines indicate the simple linear-regression fit.}
 \label{fig:linearfit}
\end{figure}

\section*{Results}
\label{sec:results}
As described in the Introduction, $\alpha$-graphyne has six possible stacking modes but only two are expected to be stable. However, DFT is unable to predict the ground state of graphyne as the energies of the two stacking modes are too close (0.6~meV/atom) for DFT to conclusively determine which is lower.~\cite{leenaerts13} Moreover,  
the energetics of the interlayer binding seems
to be highly dependent on the choice of XC functional, and on the nature of the pseudopotential used. To the best of our knowledge, the study by Leenaerts \textit{et al.} in Ref.~\cite{leenaerts13} is the only one to compute simultaneously both AB and Ab stackings within the same level of theory and using the same approximations.

In order to determine the ground state of bilayer $\alpha$-graphyne, we evaluated the binding energy per atom of a graphyne bilayer (Ab and AB stacking) using DMC by computing the energy of the system as a function of the interlayer distance $R$:
\begin{equation} 
\Delta E(R)=\frac{\left[E^{\text{bilayer}}(R)-E^{\text{bilayer}}(R=\infty)\right]}{N}.
\label{eq:morse}
\end{equation}
Here $N$ is the number of carbon atoms in a supercell and 
we took $E^{\text{bilayer}}(\infty)/2$ as the total energy of a single isolated layer.
Similar to the method used in our previous study on carbon structures,\cite{shin14} we used a linear regression fit to extrapolate the binding energies to infinite supercell size.
Figure~\ref{fig:linearfit} shows twist-averaged DMC interlayer binding energies per atom for Ab- and AB-stacked bilayer graphyne supercells of $1\times1$ ($N=16$), $2\times2$ ($N=64$), and $3\times3$ ($N=144$) as a function of $N^{-1}$. 
From the excellent fits to the thermodynamic limit, we conclude
that two-body finite size effects in QMC are effectively removed.
\begin{figure}[t]
 \includegraphics[width=6.0in]{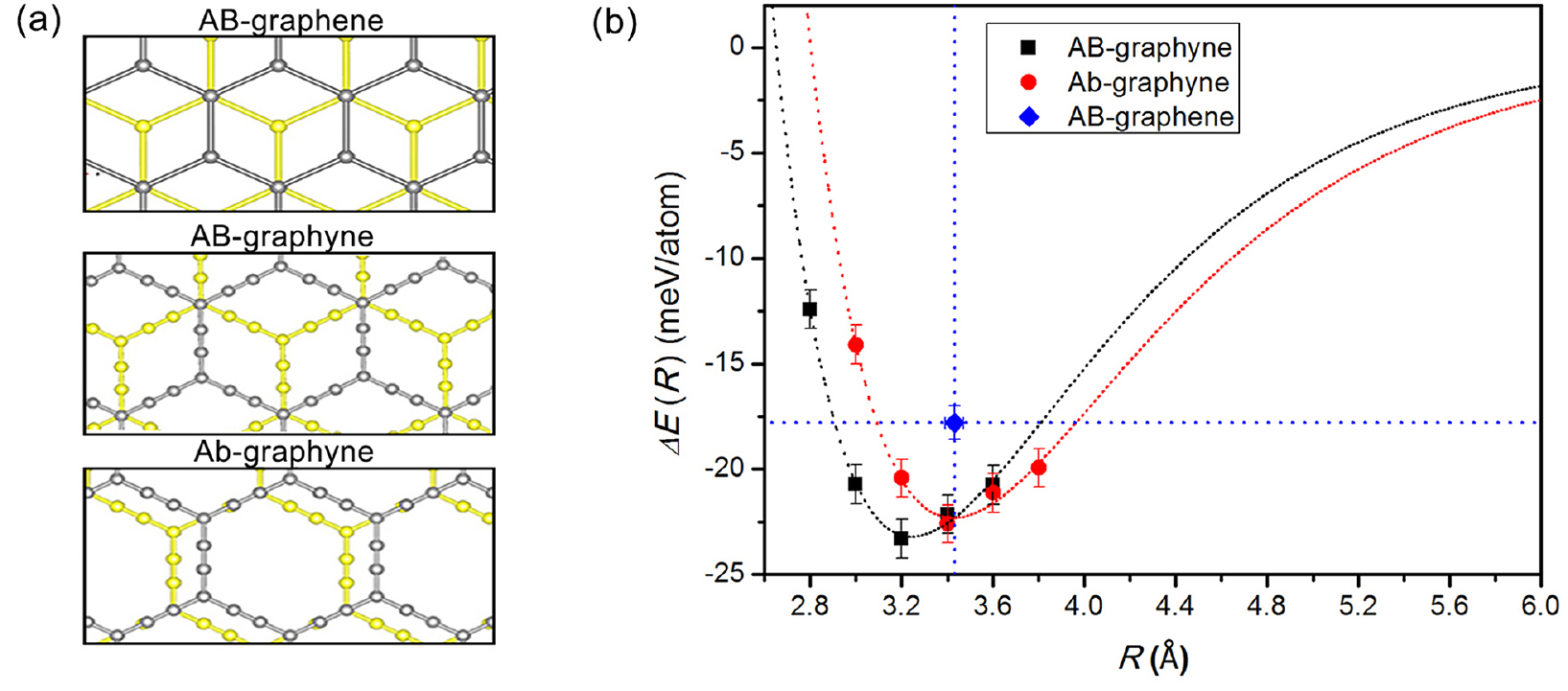}
 \caption{(a) Stacking configurations of AB stacked bilayer graphene and two stable modes (AB- and Ab-) of a bilayer $\alpha$-graphyne. The yellow and gray structures represent the low and upper layer of a bilayer, respectively. (b) DMC interlayer binding energies of AB- and Ab-stacked bilayer $\alpha$-graphynes as functions of an interlayer distance. The blue diamond symbol represents a DMC interlayer binding energy for an AB bilayer graphene at an equilibrium interlayer distance reported in Ref.~\cite{mostaani15}}
 \label{fig:qmcbinding}
\end{figure}

Figure~\ref{fig:qmcbinding} shows our DMC results for the binding energy curves of AB and Ab $\alpha$-graphyne, 
which are determined by Morse fits to the interlayer binding energies $\Delta E(R)$, and how they compare to the DMC 
binding energy of Mostaani \textit{et~al.}~\cite{mostaani15} for AB graphene.
The equilibrium interlayer distances and the interlayer binding energies
are estimated to be 3.24(1)~\AA~and 23.2(2)~meV/atom for the AB stacking mode,
and 3.43(2)~\AA~and 22.3(3)~meV/atom for the Ab mode, suggesting that the AB mode is energetically favored, albeit only by 0.9(4)~meV/atom, over the Ab mode for bilayer $\alpha$-graphyne.
The very small energy difference between these two modes suggests that it would be difficult to synthesize a pristine AB or Ab bilayer, and one can expect to have a mixture of both stacking modes at finite temperatures.   
Interestingly, the DMC interlayer binding energies of both stacking modes of graphyne are found to be noticeably larger than the corresponding DMC value of an AB-stacked bilayer graphene (the most stable stacking mode of graphene).
This suggests that the interlayer binding nature may not be purely of weak vdW form in a $sp$-$sp^2$ hybridized graphyne structure, unlike graphene.
The detailed results can be found in Table~\ref{tab:binding-QMC}.

\begin{table*}[t]
\centering
\caption{DMC equilibrium interlayer spacings $R_{0}$ (\AA) and binding energies $E_b$ (meV/atom) estimated by using the Morse function for an AB bilayer graphene, an Ab and an AB bilayer $\alpha$-graphyne. $\Delta E_{AB-Ab}$ represents the binding energy difference between AB and Ab mode of bilayer $\alpha$-graphyne.}
\label{tab:binding-QMC}
\begin{tabular}{cc|cc|cc|c}
\hline\hline
   \multicolumn{2}{c|}{graphene(AB)} & \multicolumn{2}{c|}{$\alpha$-graphyne(AB)} & \multicolumn{2}{c|}{$\alpha$-graphyne(Ab)} & \multirow{2}{*}{$\Delta E_{AB-Ab}$}  \\  \cline{1-6}
   $R_{0}$        &   $E_{b}$        &     $R_{0}$      &      $E_{b}$     &  $R_{0}$         &    $E_{b}$  &      \\ \hline
3.43(3)$^{1}$   &  17.8(3)$^{1}$   & 3.24(1) & 23.2(2) & 3.43(2) & 22.3(3) & 0.9(4) \\ \hline\hline 
\end{tabular}
\begin{flushleft}
$^1$Reference~\cite{mostaani15}.\\
\end{flushleft}
\end{table*}

\begin{figure}[t]
 \includegraphics[width=6.0in]{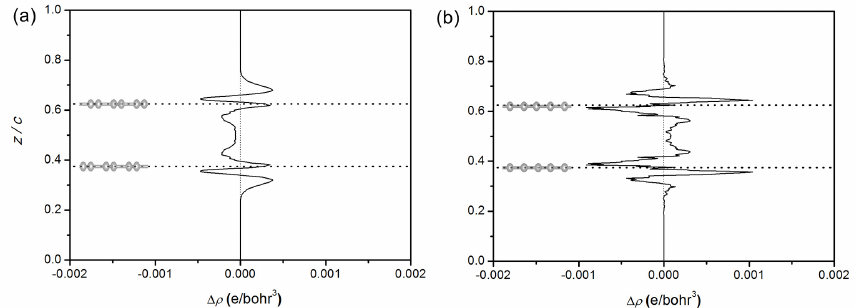}
 \caption{DMC charge density difference projected to the $z$ direction, $\Delta \rho_{\rm tot}^{z}$, on a unit cell for (a) bilayer graphene(AB) and (b) $\alpha$-graphyne(Ab) at an interlayer distance of 5~\AA; $z/c$ and the dotted lines represent
the relative positions of each upper and lower carbon layer in the bilayer unit cells.}
 \label{fig:density_QMC}
\end{figure}

The strong interlayer binding 
of graphyne over graphene cannot be attributed to vdW forces alone and must therefore be the effect of other contribution(s). Because of the weak nature of these forces, we analyzed the charge density difference projected along the $z$ axis, $\Delta \rho_{\rm tot}^{z}$, to see how charge transfer occurs between the layers. We obtain $\Delta \rho_{\rm tot}^{z}$ as
\begin{equation}
 \Delta \rho_{\rm tot}^{z}=\rho_{\rm tot}^{z}(\text{bilayer})-(\rho_{\rm tot}^{z}(\text{upper})+\rho_{\rm tot}^{z}(\text{lower})),
\end{equation} 
where $\rho_{\rm tot}^{z}(\text{bilayer})$, $\rho_{\rm tot}^{z}(\text{upper})$, and $\rho_{\rm tot}^{z}(\text{lower})$ indicate the total charge densities along the $z$ axis for a bilayer system and for the upper and lower single layers, respectively.
For more clarity, we chose to compare AB graphene to Ab graphyne as they have similar equilibrium binding distance. Figure~\ref{fig:density_QMC} shows an accumulation of charge densities around the upper and lower layer regions for both systems at a long interlayer distance of 5~\AA. 
We observe positive electron density difference at the midpoint of the interlayer region in Ab graphyne while negative density difference is seen in AB graphene at the same position.
This suggests that the larger interlayer binding energy of graphyne compared to that of graphene can be attributed to a strong contribution of covalent bonds between the layers.

\begin{figure}[t]
 \includegraphics[width=6.0in]{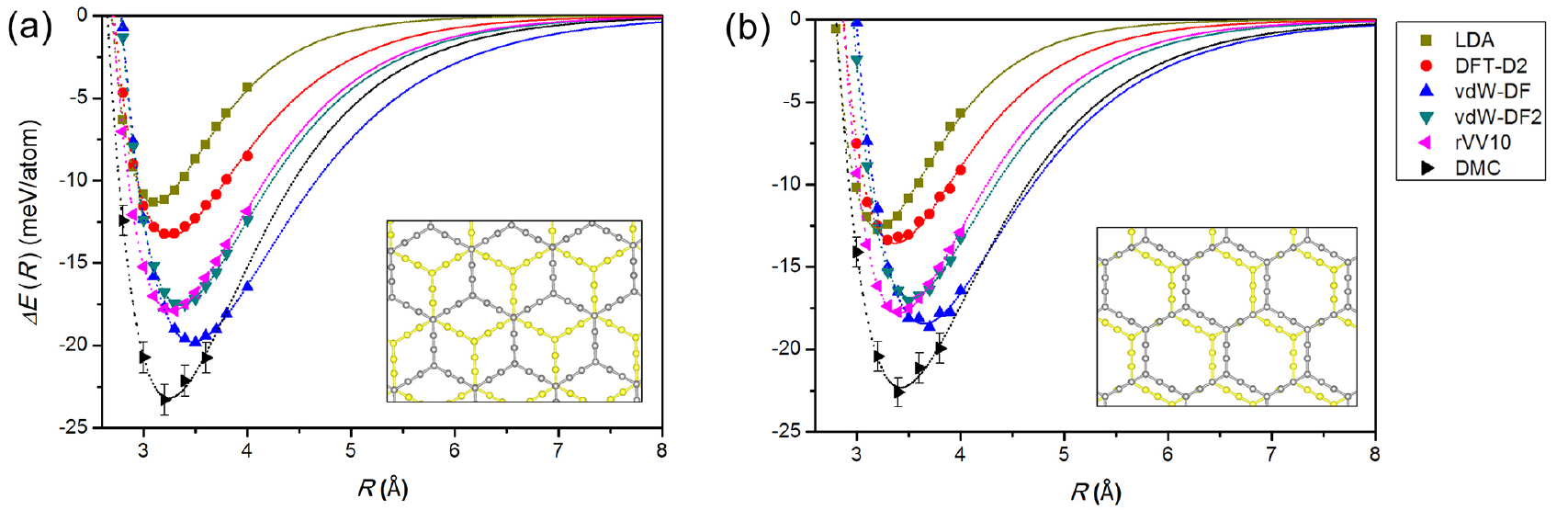}
 \caption{Interlayer binding energy for (a) AB, and (b) Ab stacking mode of a bilayer $\alpha$-graphyne using various DFT XC functionals and DMC as function of interlayer distance $R$. The dotted lines indicate the Morse function fit.}
 \label{fig:dftbinding}
\end{figure}

The DMC analysis of AB and Ab $\alpha$-graphyne and the comparison to the AB graphene allowed us to characterize the stability of the graphyne bilayer system and to conclude that their stability over graphene can be attributed to interlayer covalent bonds. DFT, in its usual forms without dispersive forces, fails to capture vdW forces but usually succeeds at describing covalent bonds. One would therefore expect DFT to reproduce the small energy difference between the two graphyne (which was shown in Ref.~\cite{leenaerts13}) stacking modes and predict graphyne to be more stable than graphene. In order to confirm this hypothesis, we compared multiple XC functionals (with and without vdW corrections).       
Although multiple DFT studies of a bilayer graphene and graphynes can be found in the literature,~\cite{lebedeva11,mapasha12,hamada10,chakarova06,leenaerts13,ozcelik13} these published results cannot be directly compared neither between them nor with our study
because of differences in pseudopotentials, reference energies, or incomplete information for a quantitative comparison. As an example, an AB bilayer graphene studied with a vdW-DF functional and a Vanderbilt ultra-soft pseudopotential as implemented in the DACAPO package finds an equilibrium interlayer distance of 3.60~\AA~and a binding energy of 45.5~meV/atom.~\cite{chakarova06} In contrast, the same system studied with the same vdW-DF functional but a projector augmented wave (PAW) pseudopotential, as implemented in the VASP package, finds an equilibrium interlayer distance of 3.35~\AA~and a binding energy of 27.1~meV/atom.~\cite{lebedeva11}   
These large variations can be attributed to strong dependence of the calculation on the pair pseudopotential and XC-correction, and to the differences in the  optimized planar geometry adopted for reference. Therefore, in order to avoid inconsistencies in our DFT calculations for comparison with our DMC results, we evaluated energies and densities of both bilayer forms of $\alpha$-graphyne using the same BFD pseudopotential and the same geometries as in our DMC study, and chose multiple vdW-corrected XC functionals as implemented in the Quantum Espresso code.  Figure~\ref{fig:dftbinding} shows the DFT interlayer binding energies of an AB- and Ab-stacked bilayer graphyne as functions of an interlayer distance using vdW-corrected XC functionals. We chose the following functionals as they use different approaches to correct for dispersion forces;
\begin{itemize}
\item DFT-D2:  a widely-used vdW correction based on an empirical dispersion term added to the total Kohn-Sham energy.~\cite{grimme04,grimme06} 
\item vdW-DF: proposed by Dion {\it et al.},~\cite{dion04} and goes beyond DFT-D2 by including non-local correlation in the XC functional.
\item vdW-DF2: an improvement over vdW-DF by replacing exchange functional in order to give more accurate description of interlayer separation than the vdW-DF.~\cite{lee10} 
\item rVV10: a more recent functional which possesses a simpler non-local correlation kernel than vdW-DF functionals.~\cite{sabatini13}
\end{itemize}

Figure~\ref{fig:dftbinding} shows that for both stacking modes, the DFT binding energy curves depend strongly on the XC functionals, even within the same type of non-local vdW corrections (vdW-DF and vdW-DF2), resulting in different equilibrium interlayer distances and different binding energies. The lack of qualitative and quantitative consensus between methods makes it of course difficult to select a \textit{best correction} scheme 
without any experimental result with which to compare. 
Therefore, using a higher level theory such as DMC for reference allows us to compare vdW-corrected functionals and will eventually help guide the choice of functionals for low-dimensional carbon allotropes. 
The analysis of Figure~\ref{fig:dftbinding} clearly shows that all the considered vdW-corrected functionals as well as a plain LDA functional significantly underestimate the interlayer binding energy of both AB and Ab $\alpha$-graphyne. This is radically different from what was observed when using these functionals to study a bilayer graphene.~\cite{mostaani15}       
Equilibrium interlayer distances $R_{0}$ and binding energies $E_{b}$ of both AB and Ab graphyne stackings as well as the AB graphene stacking, obtained using the various vdW corrected functionals, are compiled in Table~\ref{tab:binding} and compared to the DMC reference.

It is worth noting that DFT-D2 yields the weakest interlayer binding energy 
for bilayer graphyne among the vdW-corrected DFT functionals considered in this study.
This is in contrast with the previous DFT results for an AB-stacked bilayer graphene wherein the same empirical dispersion resulted in the equilibrium interlayer binding energy nearly identical to the one based on non-local vdW corrected DFT functionals within a few meV/atom binding energy differences~\cite{lebedeva11}. 
This discrepancy in the quantitative descriptions of the vdW interlayer interaction between graphyne and graphene leads us to conclude that the quantitative  contribution of each DFT vdW correction for describing the interlayer binding is not identical between $sp$-$sp^2$ hybridized carbon network and pristine $sp^2$-bonded one.
In general, all the tested vdW-corrected DFT calculations
are found to significantly underestimate the interlayer binding energies 
of bilayer graphynes, while overestimating that of a bilayer graphene. 
This gives us a hint to the explanation for the contradicting DFT and DMC results, with DFT favoring the stability of graphene over graphyne.
In the case of rVV10 functional, 
the equilibrium interlayer distances of both graphene and graphyne bilayers seem to be in excellent agreement with the corresponding DMC results. 
However, the interlayer binding energy is found to be significantly 
overestimated for graphene but underestimated for graphyne. 
On the other hand, when compared to DMC, vdW-DF and vdW-DF2 perform the best at describing binding energies despite failing at getting the right geometry. This extends to including the stability of one stacking over the other ($\Delta E_{AB-Ab}$) for $\alpha$-graphyne, which means that vdW-DF can provide both qualitatively and quantitatively accurate interlayer binding energetics for bilayer carbon-based systems.       
In conclusion, when compared to DMC results for the case of low-dimensional carbon allotropes, no vdW-correction is found to satisfy simultaneously accuracy of both interlayer distance and binding energy. However, one could imagine a scheme where the simplified non-local rVV10 functional is used to optimize geometries then the vdW-DF functional to provide the energetics of the systems.
\begin{table*}[t]
\centering
\caption{Morse fitted equilibrium interlayer distance $R_0$ (\AA) and binding energies $E_b$ (meV/atom) for an AB bilayer graphene, and Ab and AB bilayer $\alpha$-graphyne using various vdW-corrected DFT functionals. $\Delta E_{AB-Ab}$ represents the binding energy difference between AB and Ab $\alpha$-graphyne stacking.}
\label{tab:binding}
\begin{tabular}{c|cc|cc|cc|c}
\hline\hline
 \multirow{2}{*}{method}  & \multicolumn{2}{c|}{graphene(AB)} & \multicolumn{2}{c|}{$\alpha$-graphyne(AB)} & \multicolumn{2}{c|}{$\alpha$-graphyne(Ab)} & \multirow{2}{*}{$\Delta E_{AB-Ab}$}  \\ \cline{2-7}
  &    $R_{0}$       &   $E_{b}$       &     $R_{0}$      &      $E_{b}$     &  $R_{0}$         &    $E_{b}$  &      \\ \hline
LDA &    3.32    &    12.3    &     3.11      &   11.3      &    3.21     &   12.6   &  -1.3 \\                                                                                                        
DFT-D2  &  3.27      &   25.4      &  3.25      &   13.4      &  3.37    &   13.6     & -0.2  \\ 
vdW-DF &  3.62     &   24.8      &  3.47       &  19.8         &  3.64    &   18.5 &  1.3    \\ 
vdW-DF2 &  3.55      &  24.4       &  3.36      &  17.5       &  3.52    &  16.9    &  0.6    \\ 
rVV10  & 3.42      &   30.2       &   3.27      &   17.9     &   3.41     &  17.8   & 0.1  \\ 
DMC  &   3.43(3)$^{1}$   &  17.8(3)$^{1}$   & 3.24(1) & 23.2(2) & 3.43(2) & 22.3(3) & 0.9(4) \\ \hline\hline
\end{tabular}
\begin{flushleft}
$^1$Reference~\cite{mostaani15}.\\
\end{flushleft}
\end{table*}

In order to further investigate the behavior of the vdW-corrected DFT functional, 
we now focus on the effect of the corrections on the one-body Hamiltonian. The simplest correction to recover dispersion forces in DFT consists of adding a pairwise interatomic term that decays as $C_6/R^6$ to the potential obtained from DFT.\cite{grimme04,LeSar1984,Sprik-NoVdw} In the dissociation limit, or for neutral atoms with non-overlapping electron density, the leading order of the two-body dispersion contribution (dipole-dipole) to the energy corresponds to London’s formula~\cite{Heitler1927,Eisenschitz1930}
\begin{equation}
 E_{(2)}(R) = -f_{d}(R)\frac{C_{6}}{R^{6}},
\end{equation} 
where $f_{d}(R)$ represents a damping function 
(taken as $f_{d}(R)=1$ in this study).
\begin{figure}[t]
 \includegraphics[width=6.0in]{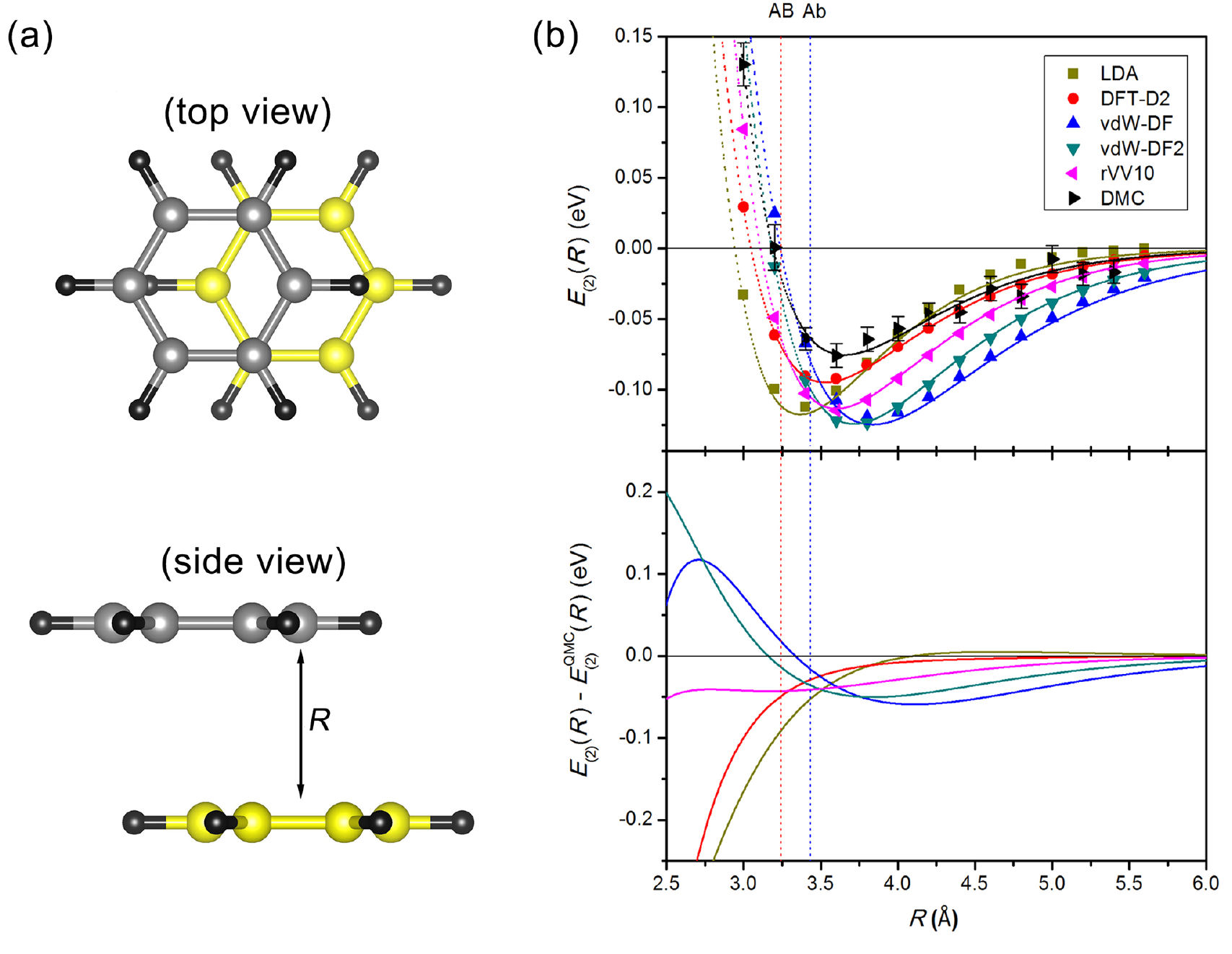}
 \caption{(a) Top-view and side-view of a parallel-displaced benzene dimer at the interplanar distance $R$, and (b) the computed two-body interaction energy (top) and its difference between DFT and DMC results (bottom) as functions of $R$. The vertical dotted lines indicate the DMC equilibrium interlayer distances for an Ab- and an AB-stacked bilayer $\alpha$-graphyne. Note that the DMC equilibrium interlayer distance for an Ab-stacked bilayer $\alpha$-graphyne was almost identical to that of an AB-stacked bilayer graphene.~\cite{mostaani15}}
\label{fig:C6}
\end{figure}
In general, the $C_6$ parameter is computed from a higher level theory, such as Coupled Cluster (CC) or Full Configuration of Interaction (FCI), applied to  the dissociation energy of a similar and simplified system.  Using DMC as a high-level theory to estimate the $C_6$ parameter has proven to be very conclusive in other vdW-dominated systems.\cite{benali14} For the purpose of the present study, we use a benzene (C$_6$H$_6$) dimer, a molecule consisting of planar carbon-based hexagonal structure and which is the closest geometry to the bilayer graphene and graphyne systems, to extract the $C_6$ parameter. Among various aligned modes of a benzene dimer, we chose a parallel-displaced (PD) one,  whose stacking mode is similar to an AB mode of a bilayer graphene or $\alpha$-graphyne (see Figure~\ref{fig:C6}a).
\begin{table}[t]
\centering
\caption{Estimated equilibrium interplanar distance $R_{0}$, two-body interaction energy $E_{(2)}(R_{0})$, and $C_6$ coefficient for a benzene dimer system with various methods based on the first-principle calculation.}
\label{tab:C6}
\begin{tabular}{cccc}
\hline\hline
  method &    $R_{0}$ (\AA)       &   $E_{(2)}(R_{0})$ (eV)       &     $C_6$ (a.u.)  \\ \hline
  LDA &    3.37    &    -0.118    &     77.71      \\ 
  DFT-D2 &    3.53    &    -0.095    &     179.77      \\ 
  vdW-DF &    3.84    &    -0.125    &     966.18      \\ 
  vdW-DF2 &    3.71    &    -0.124    &     504.43      \\ 
  rVV10 &    3.60    &    -0.113    &     271.91      \\
  DMC &    3.61(3)    &    -0.075(3)    &     145.04      \\ 
\hline\hline
\end{tabular}
\end{table}
As shown in Table~\ref{tab:C6}, all the vdW-corrected DFT two-body interaction energies are significantly overestimated compared to the DMC result, which is consistent with previous DMC calculations of a bilayer graphene.~\cite{mostaani15}
The significantly larger DFT $C_6$ values for a PD benzene system when compared to DMC, confirms that vdW-corrected DFT functionals tend to overestimate the two-body vdW dispersion forces for an $sp^{2}$-bonded hexagonal carbon network system. This is  clearly reflected in the intermolecular potential difference between DMC and DFT functionals at the equilibrium interlayer distance of AB bilayer graphene (see Figure~\ref{fig:C6}b).
Interestingly, the energy difference between rVV10 and DMC, as a function of interplanar distance is constant at short and mid-distance. 
This, along with the fact that the rVV10 functionals produce
the equilibrium interlayer distances very close to the DMC results
for $\alpha$-graphynes as well as graphene (see Table~\ref{tab:binding}),
indicates that vdW geometries within the rVV10 functional and DMC method are quantitatively consistent with one another 
for 2D carbon allotropes
including $sp$-$sp^2$ hybridized graphyne structures. 
\begin{figure}[t]
 \includegraphics[width=3.3in]{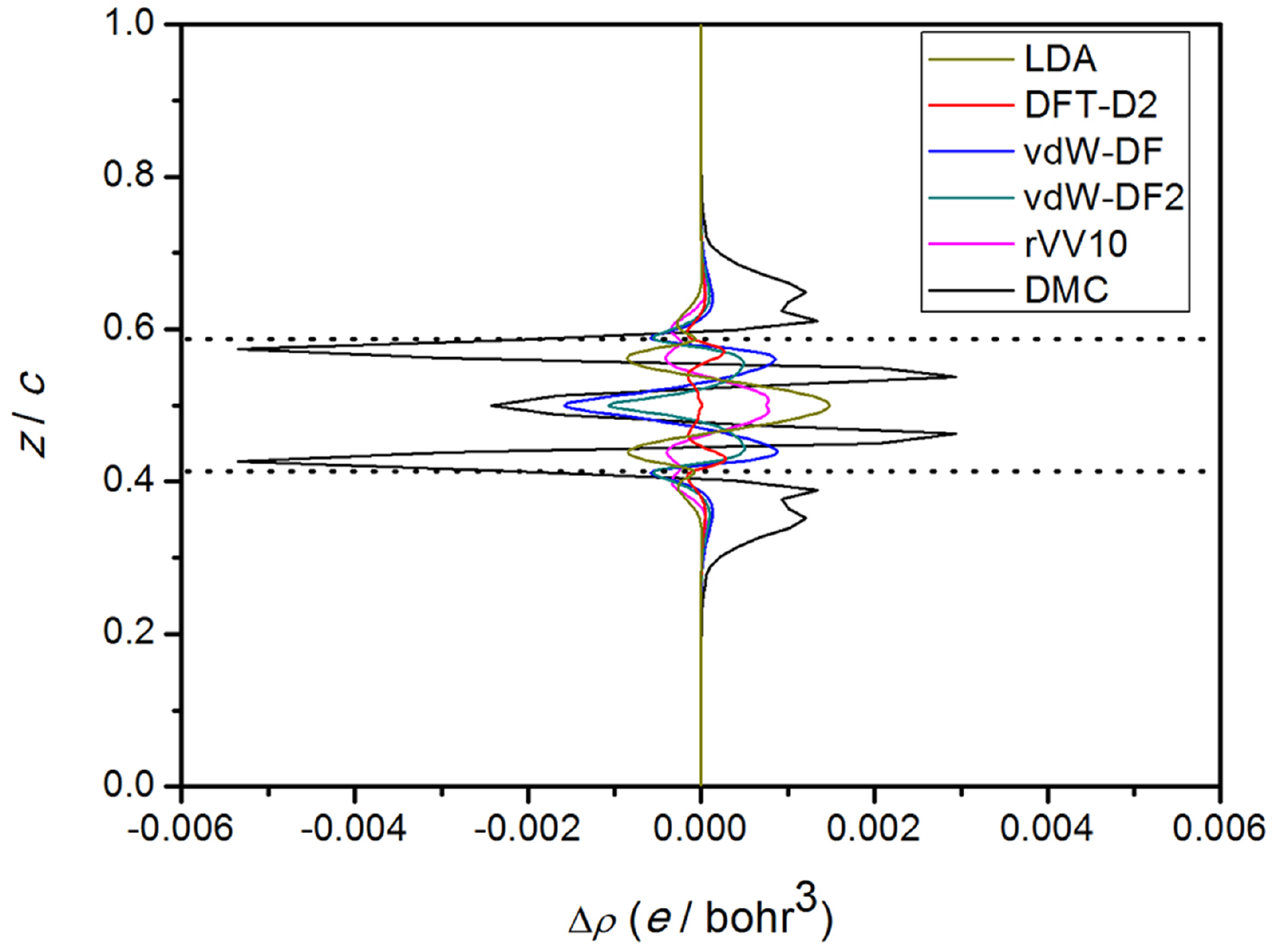}
 \caption{DFT charge density difference projected to $z$ direction on a unit cell for Ab bilayer graphyne at equilibrium interlayer distance. Note that the results for DFT-D2 are consistent with those from the PBE exchange-correlation functional.}
 \label{fig:density-DFT}
\end{figure}

As seen in the DMC analysis of the stability of the carbon allotropes, the charge density distribution conveys important information about the nature of the bonding between the graphyne/graphene layers. 
We now analyze and compare the charge density distributions from different vdW-corrected functionals and also compare them to the DMC reference in order to further assess the weaknesses and strengths of each correction scheme.  
Figure~\ref{fig:density-DFT} shows the DFT charge density differences for an Ab bilayer $\alpha$-graphyne, along with the corresponding DMC result. 
From this calculations, significantly different distributions of charge density are revealed between many-body and one-body Hamiltonians in low-dimensional $sp$-$sp^{2}$ hybridized system. Both accumulation and depletion of charge densities in the DMC result are significantly larger than those obtained from DFT XC functionals. This was also observed in other DMC studies of 2D vdW materials (black phosphorus by Shulenburger \textit{et al.}~\cite{shulenburger15}). 
Qualitatively, the vdW-DF density distributions are the closest to DMC densities for an Ab bilayer $\alpha$-graphyne at both equilibrium and long interlayer distance. This provides a good explanation for the agreements with DMC in binding energy and binding energy difference between Ab and AB bilayer graphyne (see Table~\ref{tab:binding}). 

On the other hand, while DMC shows a depletion of density in the middle of the bilayer system, rVV10, which exhibited the closest vdW geometries to the DMC results among the vdW-corrected DFT functionals, yields a completely different density distribution by generating a significantly large density accumulation in-between the layers.  
From this behavior it appears that the energetics are driven by the density while the geometry is driven by accurate dispersion forces. Therefore, there are no non-local vdW-corrected DFT functionals that satisfy both qualitative and quantitative agreements with the DMC results among the tested vdW-corrected functionals.    

\section*{Conclusions}
\label{sec:conclusions}
In summary, we have used diffusion Monte Carlo to assess the stability and nature of the interlayer binding of the two stacking modes of an $sp$-$sp^2$ hybridized graphyne,  and compared these to bilayer $sp^{2}$-bonded graphene. Because of the very small energy difference between two stable graphyne stackings, it is difficult to predict which stacking mode will be favored (if any) when temperature effects are taken into account. Most importantly and in contradiction to DFT and vdW-correct DFT predictions, DMC predicts that both stacking modes of bilayer $\alpha$-graphynes are more stable than bilayer graphene. The DMC charge density analysis of the graphene and graphyne systems attributes the higher stability of the latter to the
contribution of interlayer covalent bonds to the total energy, which are non-existent for the graphene case.

Further analysis of the DFT results shows that vdW-corrected functionals significantly underestimate the interlayer binding energies for both $\alpha$-graphyne stacking modes, while overestimating the pristine $sp^{2}$-bonded graphene. This is attributed to a different magnitude of contribution of the two-body long-range dispersion of the interlayer vdW interaction for pristine $sp^{2}$-bonded and $sp$-$sp^{2}$ hybridized bilayer carbon network.     
Among the vdW-corrected DFT functionals, 
the rVV10 electron correlation functional showed the best pair-potential $C_6$ parameter and gives an accurate description of the interlayer binding geometries for both bilayer graphene and $\alpha$-graphyne when compared to DMC. Nevertheless, inaccurate electron density distributions based on an intermolecular interaction in the interlayer region lead to a  large difference in the interlayer binding energies compared to DMC.
In contrast, vdW-DF functionals gave the closest depiction of the charge densities distribution compared to DMC, and qualitatively reproduce the DMC energetics, but fail at reproducing the $C_6$ 2-body dispersion term in a benzene dimer and in the carbon allotropes. This shows the importance of getting both the dispersion correction and the densities correct in order to obtain the right energetics and geometry when using a vdW-corrected functional.  
Our study demonstrates the stability of the AB graphyne over Ab graphyne, and in general the stability of graphyne over graphene. Moreover, it showed a direct path and a guideline to improve electron correlation functionals for explicit vdW description within the non-interacting Kohn-Sham scheme through adjustment of electron charge densities and many-body long-range dispersion.

\begin{acknowledgement}
The authors are very grateful to Luke Shulenburger (Sandia National Laboratories) and Paul Kent (Oak Ridge National Laboratory) for their help and fruitful discussions. 
The computer time for this study was provided through the Innovative and Novel Computational Impact on Theory and Experiment(INCITE) program.
This research used resources of the Argonne Leadership Computing Facility at Argonne National Laboratory, which is provided by the Office of Science of the U.S. Department of Energy (DOE) under contract DE-AC02-06CH11357. 
A.B. and H.S. were initially supported by U.S. Department of Energy,  Office  of  Science,  Basic  Energy  Sciences,  Materials  Sciences  and  Engineering Division. They, and O.H., were subsequently supported by the U.S. Department of Energy, Office of Science, Basic Energy Sciences, Materials Sciences and Engineering Division, as part of the Computational Materials.
We also acknowledge the support from the Supercomputing Center/Korea Institute of Science and Technology Information with supercomputing resources including technical support (KSC-2016-C3-001).
\end{acknowledgement}

\bibliography{bilayer}

\end{document}